# Radiative thermal switch driven by anisotropic black phosphorus plasmons

MING-JIAN HE,[1,2] HONG QI,[1,2,*] YA-TAO REN,[1,2] YI-JUN ZHAO,[1] YONG ZHANG,[1,2] JIA-DONG SHEN,[3] AND MAURO ANTEZZA[4,5]

[1]*School of Energy Science and Engineering, Harbin Institute of Technology, Harbin, 150001, China*
[2]*Key Laboratory of Aerospace Thermophysics, Ministry of Industry and Information Technology, Harbin, 150001, China*
[3] *Shanghai Aircraft Design and Research Institute, Shanghai 200000, P. R. China*
[4] *Laboratoire Charles Coulomb (L2C), UMR 5221 CNRS-Université de Montpellier, F-34095 Montpellier, France*
[5] *Institut Universitaire de France, 1 rue Descartes, F-75231 Paris, France*
*\*qihong@hit.edu.cn*

**Abstract:** Black phosphorus (BP), as a two-dimensional material, has exhibited unique optoelectronic properties due to its anisotropic plasmons. In the present work, we theoretically propose a radiative thermal switch (RTS) composed of BP gratings in the context of near-field radiative heat transfer. The simply mechanical rotation between the gratings enables considerable modulation of radiative heat flux, especially when combined with the use of non-identical parameters, i.e., filling factors and electron densities of BP. Among all the cases including asymmetric BP gratings, symmetric BP gratings, and BP films, we find that the asymmetric BP gratings possess the most excellent switching performance. The optimized switching factors can be as high as 90% with the vacuum separation $d$=50 nm and higher than 70% even in the far-field regime $d$=1 $\mu$m. The high-performance switching is basically attributed to the rotatable-tunable anisotropic BP plasmons between the asymmetric gratings. Moreover, due to the twisting principle, the RTS can work at a wide range of temperature, which has great advantage over the phase change materials-based RTS. The proposed switching scheme has great significance for the applications in optoelectronic devices and thermal circuits.

## 1. Introduction

The radiative heat transfer at the nanoscale can be enhanced by orders of magnitude [1-4] due to the contribution of photon heat tunneling, especially when the surface phonon polaritons [5-9] or surface plasmon polaritons (SPPs) [10-15] are excited. This phenomenon, known as the near-field radiative heat transfer (NFRHT), is promising for novel energy conversion technologies and nanoscale thermal management, including near-field thermostat [15], thermal routing [16], electroluminescent cooling [17], thermophotovoltaics [18], and thermal rectification [19-22], to name a few.

Due to the essential difference between electric and heat, it is common to manipulate electric currents, while to precisely control the thermal flux is still demanding. Recently, based on the unique mechanisms in the near-field regime, thermal analogues of the key building blocks in electronics have been realized, including radiative thermal diodes [23, 24], transistors [25, 26], memory elements [27], and repeater [28]. A thermal analogue to electrical switch, able to freely alternate between a low and a high thermal transport state utilizing the NFRHT, is referred to as a radiative thermal switch (RTS). Like electrical switch having ON and OFF modes, the RTS plays a crucial role in modulating the heat transfer. Specifically, the capability to switch the heat transfer has a great potential for applications in many areas [29].

Yang et al [30] proposed a RTS based on the NFRHT between phase transition materials vanadium dioxide. Nevertheless, the functions of the RTS are limited by the fixed operating temperatures. By means of electric method, radiative thermal switching was realized through changing the optical state of the gates made of electrochromic material tungsten trioxide [31]. Then the high tunability of graphene, which can be dynamically tuned by gate voltage, contributes to a radiative thermal switching scheme with graphene plasmon nanoresonators [32]. Recent theoretical works on RTS have highlighted the possibility of controlling the heat currents via entirely novel methods. Nevertheless, the above works deeply rely on the precision of the modulation manners, which are still immature technologies. An easy-to-implement modulation manner, which is reliable and can work at a wide range of temperature, is still demanding.

Black phosphours (BP), an allotrope of phosphorus, exhibits different repeating structures along the armchair and zigzag directions in each layer, hence, an in-plane anisotropy exist [33-35]. BP SPPs are electromagnetic excitations coupled to electron oscillations that propagate along the BP interface [36, 37]. Light-matter interaction with SPPs in BP nanoribbon arrays along armchair and zigzag direction have been investigated, and anisotropic plasmonic response have been observed [38]. Inspired by the unique characteristics of BP, the NFRHT between two mono/multilayer BP sheets is investigated and an enhancement of heat transfer is found, which results from a coupling of anisotropic SPPs of BP [39, 40]. Recently, by patterning BP into nanoribbon structures, the topological transition from quasi-ellipses to quasi-hyperbolas of BP SPPs has been observed [41]. The results in the above work imply that we can find two types of anisotropies in BP gratings: (1) the intrinsic anisotropy resulted from different lattice structures in the armchair and zigzag directions, and (2) the anisotropic structures caused by nanopatterning. These two types of anisotropies can couple with each other and therefore result in more complicated BP plasmons, which can be tuned by changing the electron density and structural parameters of the gratings.

In the present work, a near-field switching scheme is demonstrated based on the anisotropic black phosphorus plasmons. The paper is structured as follows. In Section 2, we introduce the physical system by presenting the configuration of the RTS, the conductivity model of BP, and the definition of heat transfer. Then in Section 3, the switching performance of the proposed RTS is demonstrated and then analyzed by the spectral radiative heat transfer coefficients, the energy transmission coefficients, the dispersion relations of surface characteristics and the reflection coefficients. The effects of the BP structures on the switching performance, including BP films, symmetric gratings, and asymmetric gratings, are compared. In addition, the switching performance is examined at different operating temperatures and with the substrate to identify the universality of the switching scheme.

## 2. Physical system

The configuration of the proposed RTS is illustrated in Fig. 1(a), which is composed of a pair of asymmetric BP gratings. The bottom and upper BP gratings are both made of periodically patterned BP nanoribbons with periodicity and width $L_1$ ($L_2$) and $L_{g,1}$ ($L_{g,2}$), respectively. The two gratings are separated by a vacuum separation denoted as $d$, and kept at temperature $T_1$ and $T_2$. It should be mentioned that the proposed RTS is at the ON and OFF modes when the two gratings are perpendicular and parallel to each other, respectively. It means that a simply mechanical rotation can switch the modes of the device. The heat transfer in the system can reach the maximum and minimum when the two gratings are perpendicular and parallel to each other, respectively. The heat transfer mechanism accounting the two cases will be explained in the following results. Recently, surface plasmons supported by two crossed layers of 2D material nanoribbons have been used to achieve dynamically tunable plasmonically induced transparency and perfect plasmonic absorptions [42-45]. Orthogonal BP nanoribbons were also used as a proof-of-concept to achieve the polarization-independent optical absorption [46]. It is confirmed that the configuration proposed in this work has been widely investigated in plasmonics and has significance for realistic devices. The detailed drawings of the BP nanoribbons in grating 1 and grating 2 are given in Fig. 1(b), which are respectively patterned along two principal lattice axes of BP, *x* (Armchair)- and *y* (Zigzag)-directions. In addition, in Fig. 1(c), the operating principle of the RTS is demonstrated by the top view of the device at the ON and OFF modes.

The conductivity of BP we used for numerical simulations is taken from Ref. [33], in which the Drude model suffice to model the conductivity of BP for photon energies of up to 0.3 eV [34]

$$\sigma_j = \frac{iD_j}{\pi(\omega + i\delta/\hbar)} \quad (1)$$

$$D_j = \pi e^2 \sum_i \frac{n_i}{m_i^j} \quad (2)$$

where $D_j$ is the Drude weight along the *j*-axis with $j \in$ (AC, ZZ). A damping constant of $\delta$ =10 meV is selected, which has been widely used in the studies of BP [39, 40, 47]. $m_i^j$ is the effective mass of the *i*-th conduction subband along the *j*-axis [48]

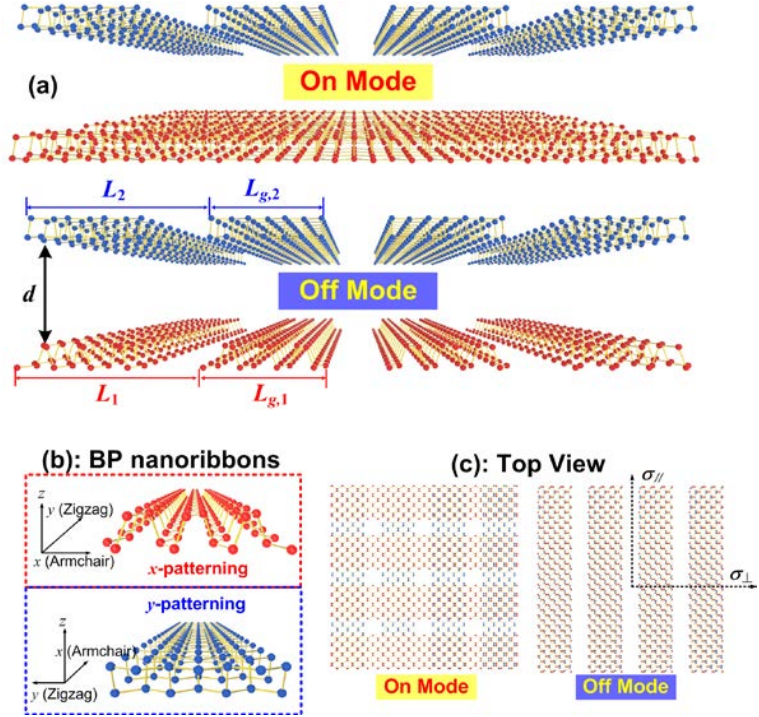

Fig. 1 (a) Schematic of the RTS composed of a pair of asymmetric BP gratings, which are made of periodically patterned BP nanoribbons with periodicity and width $L_1$ ($L_2$) and $L_{g,1}$ ($L_{g,2}$), respectively. The vacuum distance between them is denoted as *d*, and the two gratings are kept at temperature $T_1$ and $T_2$. The RTS is at the ON and OFF modes when the two gratings are perpendicular and parallel to each other, respectively. (b) Detailed drawing of the BP nanoribbons in grating 1 and 2, which are patterned along *x*(Armchair)- and *y*(Zigzag)-directions, respectively. (c) Operating principle of the RTS demonstrated by the top view.

$$m_i^{AC} = \frac{\hbar^2}{2\left[\eta_i^{AC} + \gamma_i^{AC} + \chi_i^2/(2\delta_i)\right]} \tag{3}$$

$$m_i^{ZZ} = \frac{\hbar^2}{2\left(\eta_i^{ZZ} + \gamma_i^{ZZ}\right)} \tag{4}$$

where $\eta_i^{AC}$ (in units of eV·Å$^2$)=0.364$\varsigma_i$-1.384, $\gamma_i^{AC}$ (eV·Å$^2$)=2.443$\varsigma_i$+2.035, $\chi_i$ (eV·Å$^2$)=2.071$\varsigma_i$+5.896, $\delta_i$ (eV)=0.712$\varsigma_i$+0.919, $\eta_i^{ZZ}$ (eV·Å$^2$)=2.699$\varsigma_i$+1.265, $\gamma_i^{ZZ}$ (eV·Å$^2$)=0.9765$\varsigma_i$+2.51 and $\varsigma_i$=cos[$i\pi/(N+1)$], with $N$ being the number of BP layers, which is taken as 1 in the present work. The electron density of the $i$-th conduction subband, $n_i$, is given by [35]

$$n_i = \frac{m_i^{dos}k_BT}{\pi\hbar^2}\log\left[1+\exp\left(\frac{\mu-E_i}{k_BT}\right)\right] \tag{5}$$

where $m_i^{dos} = \sqrt{m_i^{AC}m_i^{ZZ}}$ is the density of states effective mass. $E_i$(eV)=0.58$\varsigma_i$+0.505 is the minimum energy of the $i$-th conduction subband. $\mu$ is the chemical potential and the total electron density is given by $n = \sum_i n_i$.

For gratings made of anisotropic two-dimensional materials, the conductivity should be modified with the effective conductivities along ($\sigma_\parallel$) and across ($\sigma_\perp$) the main axis of the gratings, which are given by [49]

$$\sigma_\parallel = f_g\sigma_m \tag{6}$$

$$\sigma_\perp = \sigma_n\sigma_c/(f_g\sigma_c + f_c\sigma_n) \tag{7}$$

where $f_g=L_g/L$ is the BP filling factor, and $f_c=1-f_g$ is the filling factor of the free space between adjacent nanoribbons. Throughout this work, we focus on the case with the parameters selected as $L_1=L_2=10$ nm, which has been validated by the previous work in the NFRHT between BP gratings [41]. To ensure the accuracy of the calculations based on the effective medium theory, the vacuum gap distance $d$ should be several times larger than the nanoribbon periodicity. As given by Ref. [41], for $L$=10 nm, the effective medium theory predicts the real heat flux well when $d \geq 50$ nm. In the following results, the parameters are selected as $L$=10 nm with $d \geq 50$ nm, and $T$=300 K unless otherwise noted. For $x$ (Armchair)-patterning BP gratings, the conductivities obey the relations $\sigma_m=\sigma_{AC}$ and $\sigma_n=\sigma_{ZZ}$, and for $y$ (Zigzag)-patterning BP gratings, the relations are $\sigma_m=\sigma_{ZZ}$ and $\sigma_n=\sigma_{AC}$ [41]. $\sigma_c = -2i\omega\varepsilon_0\varepsilon_{eff}\frac{L}{\pi}\ln\left[\csc\left(\frac{\pi f_c}{2}\right)\right]$ [49] is an equivalent conductivity associated with the near-field coupling between adjacent nanoribbons, where $\varepsilon_{eff}$ denotes the relative permittivity of the dielectric medium surrounding BP. For two suspended BP sheets or gratings, $\varepsilon_{eff}$ is equal to 1.

In the present work, the radiative heat transfer coefficient is utilized to evaluate the NFRHT between two twisted BP gratings as [50]

$$h = \frac{\Delta\Phi}{\Delta T} = \frac{1}{8\pi^3}\int_0^\infty \hbar\omega\frac{\partial n}{\partial T}d\omega\int_0^{2\pi}\int_0^\infty \xi(\omega,\kappa,\phi)\kappa d\kappa d\phi \tag{8}$$

where $\Delta\Phi$ is the net radiative heat flux and $\Delta T = T_1 - T_2$ is the temperature difference between the two gratings. $\hbar$ is Planck's constant divided by $2\pi$ and $n=[\exp(\hbar\omega/k_B T)-1]^{-1}$ denotes the mean photon occupation number. $\xi(\omega, \kappa, \phi)$ is the energy transmission coefficient, which reads [40, 50]

$$\xi(\omega,\kappa,\phi) = \begin{cases} \text{Tr}\left[\left(\mathbf{I}-\mathbf{R}_2^\dagger\mathbf{R}_2 - \mathbf{T}_2^\dagger\mathbf{T}_2\right)\mathbf{D}_{12}\left(\mathbf{I}-\mathbf{R}_1\mathbf{R}_1^\dagger - \mathbf{T}_1\mathbf{T}_1^\dagger\right)\mathbf{D}_{12}^\dagger\right], \kappa < \kappa_0 \\ \text{Tr}\left[\left(\mathbf{R}_2^\dagger - \mathbf{R}_2\right)\mathbf{D}_{12}\left(\mathbf{R}_1 - \mathbf{R}_1^\dagger\right)\mathbf{D}_{12}^\dagger\right]e^{-2|\kappa_z|d}, \kappa > \kappa_0 \end{cases} \quad (9)$$

where $\kappa$ and $\phi$ are the surface-parallel wave vector and the azimuthal angle. $\kappa_0=\omega/c$ is the wave vector in vacuum, $\kappa_z = \sqrt{\kappa_0^2 - \kappa^2}$ is the normal component of the wave vector in vacuum. $\mathbf{D}_{12} = \left(\mathbf{I}-\mathbf{R}_1\mathbf{R}_2 e^{2i\kappa_z d}\right)^{-1}$ is the Fabry-Perot-like denominator matrix and $\mathbf{R}_j$ ($j=1, 2$) is the reflection coefficient matrix for the $j$-th BP grating, with the form

$$\mathbf{R}_j = \begin{bmatrix} r_j^{ss} & r_j^{sp} \\ r_j^{ps} & r_j^{pp} \end{bmatrix} \quad (10)$$

where the superscripts $s$ and $p$ represent the polarizations of transverse electric and transverse magnetic modes, respectively. The reflection coefficients can be referred in Refs. [51-53].

For twisted BP gratings, the conductivity tensor should be modified with the effective conductivity [54]

$$\begin{pmatrix} \sigma_{xx} & \sigma_{xy} \\ \sigma_{yx} & \sigma_{yy} \end{pmatrix} = \begin{pmatrix} \sigma_\| \cos^2\phi_j + \sigma_\perp \sin^2\phi_j & (\sigma_\| - \sigma_\perp)\sin 2\phi_j / 2 \\ (\sigma_\| - \sigma_\perp)\sin 2\phi_j / 2 & \sigma_\| \sin^2\phi_j + \sigma_\perp \cos^2\phi_j \end{pmatrix} \quad (11)$$

$\phi_j$ in Eq. (11) denotes the azimuthal angle of the main axis for grating $j$ ($j=1, 2$). When the gratings are aligned, the relationship of azimuthal angles for the two gratings is $\phi_2= -\phi_1$. If a twisted angle $\theta$ exists between the main axes of the two gratings, then $\phi_2=-\phi_1+\theta$ [50]. In view of Eq. (11), integration over azimuthal angles is necessary for the energy transmission coefficients because Fresnel's coefficients and conductivity of BP gratings depend on $\phi$. Hence, the energy transmission coefficients considering the integration over azimuthal angles are given as

$$\xi(\omega,\kappa) = \frac{1}{2\pi}\int_0^{2\pi} \xi(\omega,\kappa,\phi)d\phi \quad (12)$$

### 3. Results and discussion

To measure the performance of the RTS, the switching factor is defined as follows

$$\eta = \left(1 - h_{\text{off}} / h_{\text{on}}\right) \times 100\% \quad (13)$$

where $h_{\text{off}}$ and $h_{\text{on}}$ denotes the heat transfer coefficients of the RTS at the OFF and ON mode, respectively. As is demonstrated in Fig. 1, when the vacuum distance $d$ is settled, the parameters that can influence the NFRHT are the BP filling factors ($f_{g,1}/f_{g,2}$) and the electron densities ($n_1/n_2$) of the two BP gratings. To obtain the optimized performance of the RTS, the $\eta$ should be evaluated at different combinations of ($f_{g,1}$, $f_{g,2}$, $n_1$, $n_2$), which is a time-consuming work. To solve this problem, a stochastic particle swarm optimizer (SPSO) algorithm [55], which can guarantee the convergence of the global optimization solution, is adopted to optimize the performance of the RTS. Compared with standard particle swarm optimizer algorithm, SPSO eliminates the historical velocity term which makes the particle lose velocity memory to decrease the global searching ability. But it guarantees at every generation one particle stops evolution for locating at the best position. Due to the grating configuration and the scope of electron density in Drude model, the lower and upper boundary of the parameters are set as $f_g \in [0.1, 0.95]$ and $n \in [1\times10^{12} cm^{-2}, 5\times10^{13} cm^{-2}]$, respectively. In the primary optimizing process, the optimized results show that at different $d$, the electron densities of the two BP gratings have two different trends. The electron density $n_1$ of grating 1, which is $x$-patterning, tends to be the upper boundary of the scope $[1\times10^{12} cm^{-2}, 5\times10^{13} cm^{-2}]$. Conversely, the electron density $n_2$ of grating 2, which is $y$-patterning, is nearly at the lower boundary $1\times10^{12} cm^{-2}$ of the scope. Inspired by the phenomenon, in the subsequent optimizing process, the electron densities of BP grating 1 and grating 2 are set as $5\times10^{13} cm^{-2}$ and $1\times10^{12} cm^{-2}$, respectively. Thus the four-parameter optimizing problem reduces to a two-parameter optimization, which is easy to obtain the global optimization solution.

In Fig. 2(a), the optimized switching performance of the RTS, is demonstrated by the switching factors at different vacuum separation $d$. The results show that $\eta$ reaches nearly 90% at the vacuum separation $d=50$ nm. As $d$ increases, the switching performance shows a monotonically decrease due to the attenuation effect of BP SPPs at the long distance. However, the $\eta$ is still larger than 70% even at the large distance $d=1$ $\mu$m. The heat transfer coefficients at the ON and OFF modes are also demonstrated in Fig. 2(a) to reveal the trend of the NFRHT in the system. An interesting phenomenon can be observed that near $d=300\sim400$ nm, a rapid growth of the heat transfer coefficients emerges. It implies that the RTS can still maintain a relatively strong heat transport in the far-field, and thus the practicality of the RTS is guaranteed. The optimized filling factors of BP gratings for the gratings 1 and grating 2 are

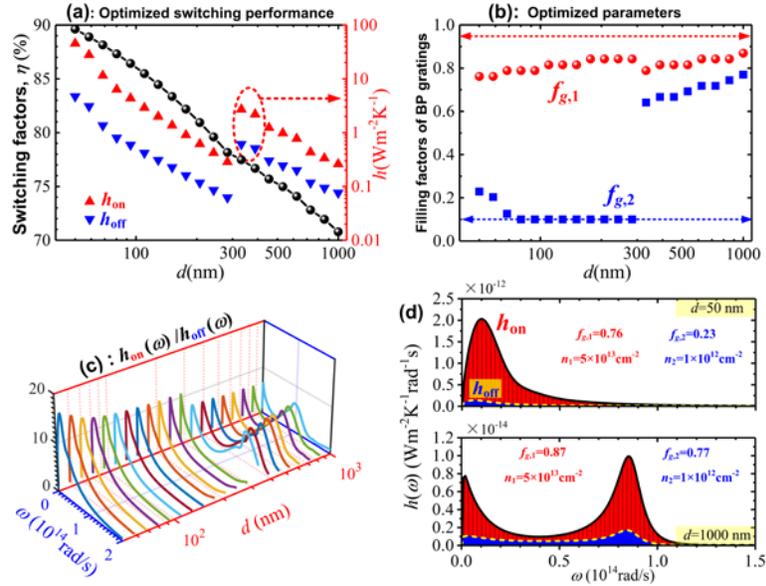

Fig. 2 (a) The optimized switching performance of the RTS at different vacuum separation $d$, demonstrated by the switching factors $\eta=(1-h_{off}/h_{on})\times100\%$ and the heat transfer coefficients at the ON and OFF modes. (b) The optimized filling factors of grating 1 and grating 2, with the optimized electron densities fixed at $n_1=5\times10^{13}$ cm$^{-2}$ and $n_2=1\times10^{12}$ cm$^{-2}$, respectively. (c) The ratio of spectral heat transfer coefficients at the ON mode $h_{on}(\omega)$ to that of the OFF mode $h_{off}(\omega)$. (d) For $d=50$ nm and 1000 nm, the spectral heat transfer coefficients at the ON mode $h_{on}(\omega)$ and OFF mode $h_{off}(\omega)$. The optimized parameters are indicated in the figure.

given in Fig. 2(b). The lower and upper boundaries of the parameters selected in the optimization are labeled with dashed lines in blue and red, respectively. The filling factors of x-patterning grating $f_{g,1}$ varies around 0.8 at different $d$, while $f_{g,2}$ has an entirely different trend with $d$. Around $d$=300~400 nm, a sudden change of $f_{g,2}$ emerges and it jumps to 0.64 at $d$=331 nm. Then as $d$ increases, $f_{g,1}$ and $f_{g,2}$ have the same trend with $d$ and they both rise. Thus, the sudden change of heat transfer coefficients observed in Fig. 2(a) can be attributed to the variety of the BP filling factors. To explore the NFRHT in the system, in Fig. 2(c), the ratio of spectral heat transfer coefficients at the ON mode to that of the OFF mode is demonstrated at different $d$. It shows that at $d$=50 nm, the ratio $h_{on}(\omega)/h_{off}(\omega)$ can reach as high as 16. Around $d$=300~400 nm, the spectral ratio also reveals a sudden change of NFRHT, i.e., the single-peak transforms to double-peak. To specify the two different behaviors in detail, in Fig. 2(d), the spectral heat transfer coefficients are plotted for $d$=50 nm and 1000 nm, respectively. The results show that the NFRHT in the system reveals two different mechanisms in the near- and far-field regime.

To explore the underlying mechanism of the switching performance, the energy transmission coefficients $\text{Log}_{10}(\xi(\omega,\kappa))$ at the ON and OFF modes are demonstrated in Figs. 3(a) and 3(b) for $d$=50 nm, respectively. The results obviously show that the BP SPPs decay dramatically at all frequencies when the two gratings are parallel to each other. In Figs. 3(c) and 3(d), the evolution of the corresponding results $\text{Log}_{10}(\xi(\omega,\kappa_x, \kappa_y)\cdot\kappa)$ with frequencies are given. We show that the BP SPPs reveal totally different regimes at the ON and OFF modes. The most important point is that at all frequencies $\omega$=1×10$^{13}$ rad/s~4×10$^{13}$ rad/s, the $\xi_{on}(\omega,\kappa_x, \kappa_y)\cdot\kappa$ are larger than $\xi_{off}(\omega,\kappa_x, \kappa_y)\cdot\kappa$. The dispersion relations of BP SPPs are added in Figs. 3(c) and 3(d) with dashed lines to assist to analyze the evolution of the energy transmission coefficients. We find that the dispersion relations of BP SPPs corresponding to x-patterning, are much narrower than those of y-patterning grating. As $\omega$ increases to be higher than 1×10$^{13}$ rad/s, the dispersion relations y-patterning cannot be found in the figure any more.

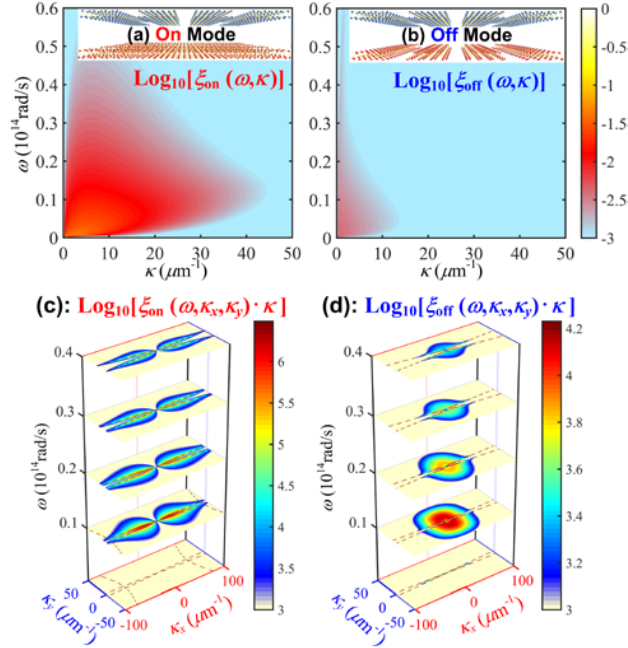

Fig. 3 At $d$=50 nm, the common logarithm of energy transmission coefficients $\text{Log}_{10}(\xi(\omega,\kappa))$ at the (a) ON and (b) OFF modes. The corresponding results of $\text{Log}_{10}[\xi(\omega,\kappa_x, \kappa_y)\cdot\kappa]$ at the (c) ON and (d) OFF modes at different frequencies. The dispersion relations of the surface characteristics for the two BP gratings are indicated with dashed lines.

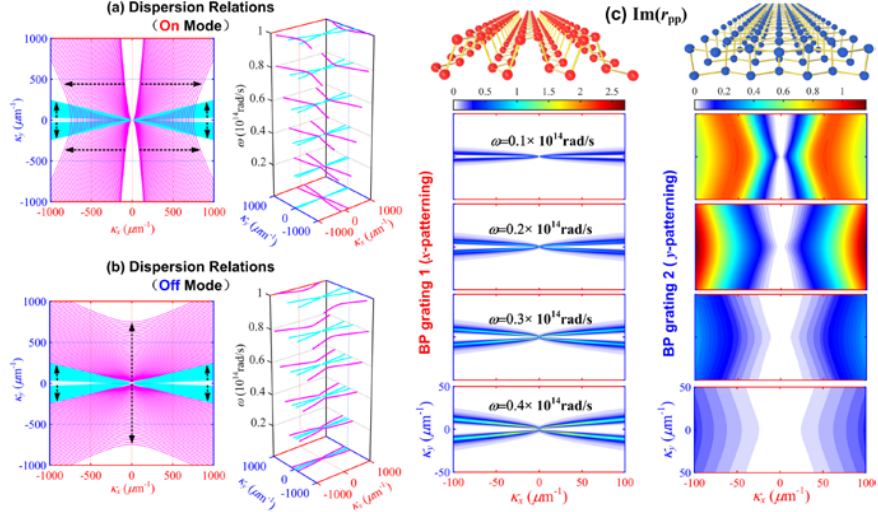

Fig. 4 The evolution of dispersion relations with frequencies for the two BP gratings at (a) ON mode and (b) OFF modes. The curves in cyan and magenta denote the dispersion relations for BP SPPs excited by *x*-patterning and *y*-patterning gratings, respectively. The arrows in the panels of the left column point to the directions where the dispersion relations tend towards when $\omega$ increases. (c) The imaginary parts of *p*-polarized reflection coefficients $r_{pp}$ for the two BP gratings at different frequencies.

By tuning the electron density of BP or changing the filling factors of the gratings, the topological transitions of BP SPPs can be adjusted [41]. To characterize the two different modes of the RTS and distinguish the different surface characteristics of the two gratings, the evolution of the dispersion relations with frequencies for the two BP gratings are illustrated in Figs. 4(a) and 4(b). The curves in cyan and magenta denote the dispersion relations for BP SPPs excited by *x*-patterning and *y*-patterning gratings, respectively. The left panels in Figs. 4(a) and 4(b) demonstrate the dispersion relations at different frequencies in a two-dimensional space $\kappa_x$-$\kappa_y$. The arrows point to the directions where the dispersion relations tend to when $\omega$ increases. The dispersion relations of the two gratings are both hyperbolic in the frequency region of interest. Furthermore, they both expand in the $\kappa_x$-$\kappa_y$ space as $\omega$ increases. However, the hyperbolic curves corresponding to the *x*-patterning gratings are always much narrower than those of *y*-patterning gratings. At the ON mode, the two couples of hyperbolic curves intersect at all frequencies due to the nature of hyperbolic curves. However, at the OFF mode, when the two gratings have the parallel main axes, the two couples of hyperbolic curves cannot intersect anymore. An interesting phenomenon can be found in the left panel of Fig. 4(b) that, the cyan hyperbolic curves fill the region between the magenta curves, and thus they have no intersection at all. It is mainly attributed to the great mismatch between the two totally different surface characteristics, which is resulted from the different patterning manners.

In Fig. 4(c), the imaginary parts of *p*-polarized reflection coefficients $r_{pp}$ are demonstrated. Specifically, $r_{pp}$ plays a crucial role in the NFRHT. The results in the left column and right column represent the Im($r_{pp}$) of *x*-patterning and *y*-patterning gratings, respectively. An obvious mismatch of reflection coefficients is observed between the two gratings. For the *x*-patterning grating, we show that the Im($r_{pp}$) is larger than those of *y*-patterning grating. In addition, the hyperbolic $r_{pp}$ is much narrower than that of *y*-patterning grating and is restricted in narrow $\kappa_y$ space. As $\omega$ increases, the Im($r_{pp}$) enhances and expands to higher $\kappa_y$. For *y*-patterning grating, the spreading effect of Im($r_{pp}$) is much more obvious than that of *x*-patterning grating. In the previous studies between two gratings with identical parameters [4, 52], the heat flux reaches the maximum and the minimum when the gratings are parallel and

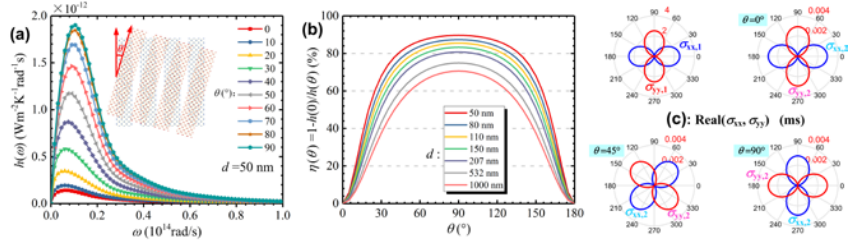

Fig. 5 (a) At $d$=50 nm, the spectral heat transfer coefficients for different twisted angles $\theta$ between the two BP gratings, which is illustrated by the inset. (b) The switching factors $\eta(\theta)$=1-$h$(0)/$h(\theta)$ as the function of twisted angle. (c) At $\omega$=1×10$^{13}$ rad/s, the real parts of $\sigma_{xx}$ and $\sigma_{yy}$ for grating 1 (the top left panel), and grating 2 at $\theta$=0°(the top right panel), 45° (the bottom left panel) and 90° (the bottom right panel).

perpendicular to each other. When the two gratings are parallel, the plasmons are strongly excited due to the match between the two interfaces. However, the modes dramatically decay when the two gratings are perpendicular to each other, for the reason that the couple modes are broken by the mismatch. As for the proposed RTS in the present work, the mechanism is totally contrary to that of the previous studies. The strong mismatch in the reflection coefficients between the two gratings makes it possible to realize broken coupling of BP SPPs between the two interfaces. For the reason that SPPs can only be excited near the region where the dispersion relations of the interfaces intersect, the mismatched surface characteristics break the coupling when the dispersion relations cannot meet at $\theta$=0°, as demonstrated in Fig. 4(b).

To reveal the rotation effect on the switching performance, in Fig. 5(a), the spectral heat transfer coefficients are plotted for different twisted angles $\theta$. As $\theta$ increases, the heat transfer enhances and reaches the maximum at $\theta$=90°. The enhancing trend of NFRHT with twisted angle, has never been observed in the previous studies considering twisting between two anisotropic materials [4, 56-59]. In Fig. 5(b), the switching factors $\eta(\theta)$=1-$h$(0)/$h(\theta)$ are plotted for different $d$. $\eta(\theta)$ increases rapidly with $\theta$ and nearly more than 2/3 of the maximum switching factor is achieved at $\theta$=30°. It implies that the RTS slightly relies on the precision of the mechanical rotation. At $\omega$=1×10$^{13}$ rad/s, the real parts of $\sigma_{xx}$ and $\sigma_{yy}$ for grating 1 and grating 2 at $\theta$=0°, 45° and 90° are given in Fig. 5(c). The anisotropy of $\sigma_{xx}$ and $\sigma_{yy}$ are obviously demonstrated and the different magnitudes are also revealed. As $\theta$ increases from 0° to 90°, both the $\sigma_{xx}$ and $\sigma_{yy}$ rotate in the polar coordinates. When the gratings are perpendicular, i.e., $\theta$=90°, an alignment is observed between $\sigma_{xx,1}$ and $\sigma_{yy,2}$, and between $\sigma_{xx,2}$ and $\sigma_{yy,1}$. Together with the results in Fig. 4, the rotation effect on the switching performance can be well explained.

To highlight the superiority of the asymmetric structure, the switching factors for different cases are demonstrated in Fig. 6(a): the optimized asymmetric BP gratings, symmetric BP gratings, BP films at different electron densities. The symmetric gratings are composed of two identical gratings, which are both in $x$-patterning or $y$-patterning, respectively. For symmetric BP gratings, the heat transfer reaches the maximum and minimum when the two gratings are parallel and perpendicular to each other, respectively. For BP films, the maximum and minimum NFRHT are achieved when the two films have the same and orthogonal directions of lattice axes. The mechanism obeys the same relation with the previous studies about NFRHT between anisotropic materials [4, 52]. Therefore, the switching factors for the two cases are defined as $\eta$=1-$h$(90°)/$h$(0). The SPSO method is utilized to optimize the switching performance of the symmetric gratings at different combinations of ($f_{g,1}$, $f_{g,2}$, $n_1$, $n_2$). The optimized results show that when the electron densities of BP and the filling factors of the bottom and upper BP gratings are identical, the switching

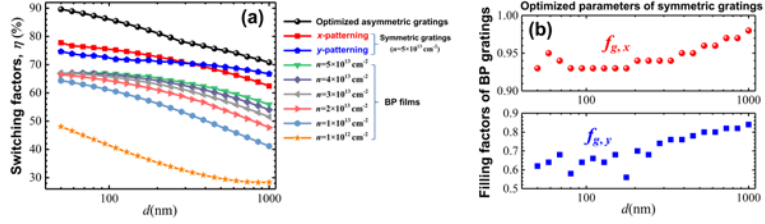

Fig. 6 (a) The switching factors for different cases: the optimized asymmetric BP gratings, symmetric BP gratings in *x*-patterning and *y*-patterning, BP films at different electron densities. For symmetric BP gratings and BP films, the switching factor is defined as $\eta=1-h(90°)/h(0)$, for the reason that the heat transfer reaches the maximum and minimum at $\theta=0°$ and $\theta=90°$, respectively. The trend is inverse to that of asymmetric BP gratings. (b) The optimized filling factors of symmetric BP gratings in *x*-patterning and *y*-patterning, respectively.

performance is strongest. In addition, the best performance is achieved with the maximum electron densities $n_1=n_2=5\times10^{13}$ cm$^{-2}$. We show that by means of asymmetric configurations, in the near-field regime, the switching factors can be increased by about 15% than the symmetric gratings and about 25% than the BP films. In the far-field regime $d=1$ μm, the enhancement of the switching performance caused by the asymmetric configuration is still strong. In Fig. 6(b), the optimized filling factors of symmetric BP gratings are given. The filling factors of *x*-patterning gratings, $f_{g,x}$, varies in a small range 0.9~1. Nevertheless, the range is much larger for $f_{g,y}$, which is nearly 0.5~0.9. The differences in optimized parameters and switching performance between *x*-patterning and *y*-patterning symmetric gratings, reveal the strong mismatch induced by different patterning manners. This can also contribute to the explanation of the excellent performance of the asymmetric gratings. In Fig. 6(a), the results of BP films are always worse than those of gratings, no matter in any grating structures. It is mainly attributed to the weak anisotropy of BP films, which can only be induced by the armchair and zigzag lattices in BP. As discussed above, the grating configuration endows the system with another manner to enhance the anisotropy in the in-plane directions. It is just the strong and twofold anisotropy that results in the high performance of RTS.

To examine the universality of the proposed RTS, the switching factors at different temperatures are demonstrated in Fig. 7(a) when the parameters of the RTS are same as those of Fig. 2. As the temperature of BP increases, the switching factors slightly decay in the near field regime, while in the far-field regime they are enhanced. The different trends of switching factors with $T$ are mainly attributed to the different mechanisms observed in Fig. 2 in different ranges of $d$. Although the switching factors vary with the operating temperatures, they still stay at relatively high values at different $d$. We show that the RTS can work at a wide range of temperature due to the twisting principle, which has great advantage over the phase change materials-based RTS.

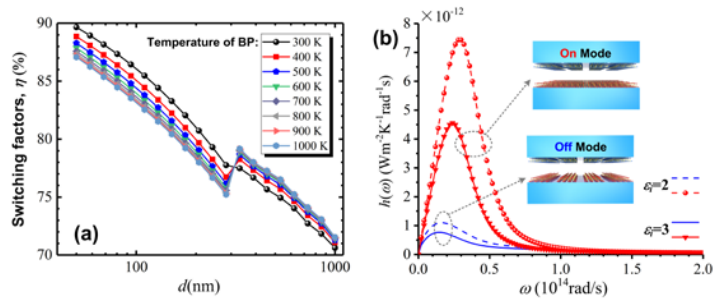

Fig. 7 The universality of the RTS: (a) the switching factors at different temperatures; (b) the spectral heat transfer coefficients at the ON and OFF modes when the BP gratings are coated on substrate with permittivity $\varepsilon_i=2$ and $\varepsilon_i=3$.

Based on the current nanotechnology, it is challenging to implement such a suspended grating system, especially with the tiny vacuum layer and the parallelism of the two gratings. However, when BP is coated on the substrate, the switching can be easily realized by simply mechanical rotation between the upper and the bottom objects. In addition, two-dimensional materials are always coated on the substrate in the actual devices. In Fig 7(b), the switching performance is demonstrated by the spectral heat transfer coefficients at the ON and OFF modes when the BP gratings are coated on substrate with permittivity $\varepsilon_i=2$ and $\varepsilon_i=3$ at $d=50$ nm. The SPSO method is utilized to obtain the optimized switching performance, which are $\eta=78.58\%$ and $\eta=72.39\%$ for the permittivity $\varepsilon_i=2$ and $\varepsilon_i=3$, respectively. The switching performance decays after the substrate is added. It can be well explained that, the adding of substrate weakens the dominated role that BP grating plays in the system. Although the switching performance is weakened, the results obtained here still have great significance in the laboratory physics when two-dimensional materials are always attached to the substrate. We expect that the performance of the RTS can be improved when the substrate can couple with the coated BP gratings. However, like silica [15] or silicon carbide [3, 4], which can strongly interact with the coated films with surface phonon polaritons, will introduce totally different physical regimes in the NFRHT. Therefore, in the present work, we restrict our attention only on the BP gratings. The results obtained in the present work are not influenced by the other materials and thus the pure effects of BP can be understood well, which paves a way for the BP-based technology.

### 4. Conclusion

In the present work, a radiative thermal switching scheme is theoretically proposed based on anisotropic black phosphorus plasmons. The RTS is composed of two asymmetric BP gratings, which are made of BP nanoribbons periodically patterned along different lattice axes. A simply mechanical rotation can switch the modes of the device, i.e., the NFRHT reaches the maximum and minimum when the two gratings are orthogonal and parallel, respectively. The enhancing trend of NFRHT with twisted angle, has never been observed in the previous studies considering rotation between two anisotropic materials. In addition, the switching function slightly relies on the precision of the mechanical rotation. By tuning the parameters of the two BP gratings, the switching performance of the proposed RTS is optimized, which can be as high as 90% with the vacuum separation $d=50$ nm and higher than 70% even in the far-field regime. It is found that the mismatch between the two interfaces is the key to realize the switching function, which benefits from the twofold anisotropy in BP gratings. Among all the cases including asymmetric BP gratings, symmetric BP gratings, and BP films, we find that asymmetric BP gratings possess the most excellent switching ability. Moreover, due to the twisting principle of the RTS, the performance can be slightly influenced by the operating temperature, which has great advantage over the phase change materials-based RTS.

Due to the limitations of the effective medium theory used in the present work, the switching performance cannot be precisely simulated below $d=50$ nm. As a result of the monotonicity of switching factors with $d$, we expect a better switching performance in the ultra near-field regime $d<50$ nm. In the future research, one can explore the switching performance of the proposed RTS in extremely small separation via accurate scattering theory based on the rigorous coupled-wave analysis method [41] or the full many-body radiative heat transfer theory [60].

### Funding

This work was supported by the National Natural Science Foundation of China (No. 51976044, 51806047, 51706053).

# Acknowledgments

The authors acknowledge support from Heilongjiang Touyan Innovation Team Program. M. A. acknowledges support from the Institute Universitaire de France, Paris, France (UE).

# Disclosures

The authors declare no conflicts of interest.